\documentstyle[12pt,emulateapj]{article}

\input{epsf}

\journalid{}{}
\articleid{}{}
\begin{document}
\def\lax    {\ifmmode{_<\atop^{\sim}}\else{${_<\atop^{\sim}}$}\fi}
\def\gax    {\ifmmode{_>\atop^{\sim}}\else{${_>\atop^{\sim}}$}\fi}
\def\gtorder{\mathrel{\raise.3ex\hbox{$>$}\mkern-14mu
             \lower0.6ex\hbox{$\sim$}}}
\def\ltorder{\mathrel{\raise.3ex\hbox{$<$}\mkern-14mu
             \lower0.6ex\hbox{$\sim$}}}
 
\long\def\***#1{{\sc #1}}
 
\title{The discovery of 2.78 hour periodic modulation of the X-ray flux from 
globular cluster source Bo 158 in M31.}

\author{Sergey P. Trudolyubov\altaffilmark{1}, 
Konstantin N. Borozdin\altaffilmark{1}, William C. Priedhorsky}
\affil{Los Alamos National Laboratory, Los Alamos, NM 87545}

\altaffiltext{1}{{\it Also:} Space Research Institute, 
Russian Academy of Sciences, Profsoyuznaya 84/32, Moscow 117810, Russia}

\author{Julian P. Osborne, Michael G. Watson}
\affil{Department of Physics $\&$ Astronomy, University of Leicester, 
Leicester LE1 7RH, UK}

\author{Keith O. Mason}
\affil{Mullard Space Science Laboratory, University College London, 
Holmbury St. Mary, Dorking, Surrey, UK}

\author{France A. Cordova}
\affil{University of California, Riverside}

\begin{abstract}
We report the discovery of periodic intensity dips in the X-ray source
XMMU J004314.1+410724, in the globular cluster Bo158 in M31. The X-ray 
flux was modulated by $\sim 83\%$ at a period of 2.78 hr (10017 s) in 
an XMM-Newton observation taken 2002 Jan 6-7. The X-ray intensity dips 
show no energy dependence. We detected weaker dips with the same period 
in observations taken 2000 June 25 ({\it XMM-Newton}) and 1991 June 26 
({\it ROSAT}/PSPC). The amplitude of the modulation has been found to be 
anticorrelated with source X-ray flux: it becomes lower when the source 
intensity rises. The energy spectrum of Bo158 was stable from 
observation to observation, with a characteristic cutoff at $\sim 4 - 6$ 
keV. The photo-electric absorption was consistent with the Galactic 
foreground value. No significant spectral changes were seen in the course 
of the dips. If the 2.78 hr cycle is the binary period of Bo158 the system 
is highly compact, with a binary separation of $\sim 10^{11}$ cm. The 
association of the source with a globular cluster, together with spectral 
parameters consistent with Galactic neutron star sources, suggests that 
X-rays are emitted by an accreting neutron star. The properties of Bo 158 
are somewhat reminiscent of the Galactic X-ray sources exhibiting a dip-like 
modulations. We discuss two possible mechanisms explaining the 
energy-independent modulation observed in Bo 158: i) the obscuration of the 
central source by highly ionized material that scatters X-rays out of the 
line of sight; ii) partial covering of an extended source by an opaque 
absorber which occults varying fractions of the source.
\\
\\
{\em Subject headings:} galaxies: individual (M31) --- stars: individual 
(Bo158) --- galaxies: star clusters --- X-rays: galaxies --- X-rays: stars
\end{abstract}

\section{INTRODUCTION}
The X-ray source XMMU J004314.1+410724 was discovered in M31 by the 
{\em Einstein} observatory (source $\#85$ in \cite{TF91}) and was detected 
in subsequent observations with {\em ROSAT} (\cite{Primini93,Supper01}), 
{\em XMM-Newton} (\cite{Shirey01}) and {\em Chandra} (\cite{DiS01}). Based 
on the {\em Chandra} aspect solution, which is currently limited by 
systematics to $\sim0.6\arcsec$ accuracy, the source location is 
$\alpha = 00^{h} 43^{m} 14.42^{s}$, $\delta = 41\arcdeg 07\arcmin 26.3\arcsec$ (2000 equinox) (\cite{DiS01})(Fig. \ref{image_mos}). This position of the 
source is consistent with the optically identified globular cluster candidate 
Bo 158 (source $\#158$ in Table IV of \cite{Battistini87}) (we will use 
the designation ``Bo 158'' as a source name throughout this Letter). 

The previous observations were too insensitive for detailed study of 
individual sources in M31. The large collecting area and bandpass of 
{\it XMM-Newton} allow us to study the short-term variability and 
spectral properties of these sources. In this Letter we report the 
discovery of periodic X-ray modulation of the lightcurve of Bo 158 
and report its spectra.

\section{OBSERVATIONS AND DATA ANALYSIS}
\begin{figure}
\epsfxsize=18.0cm
\epsffile{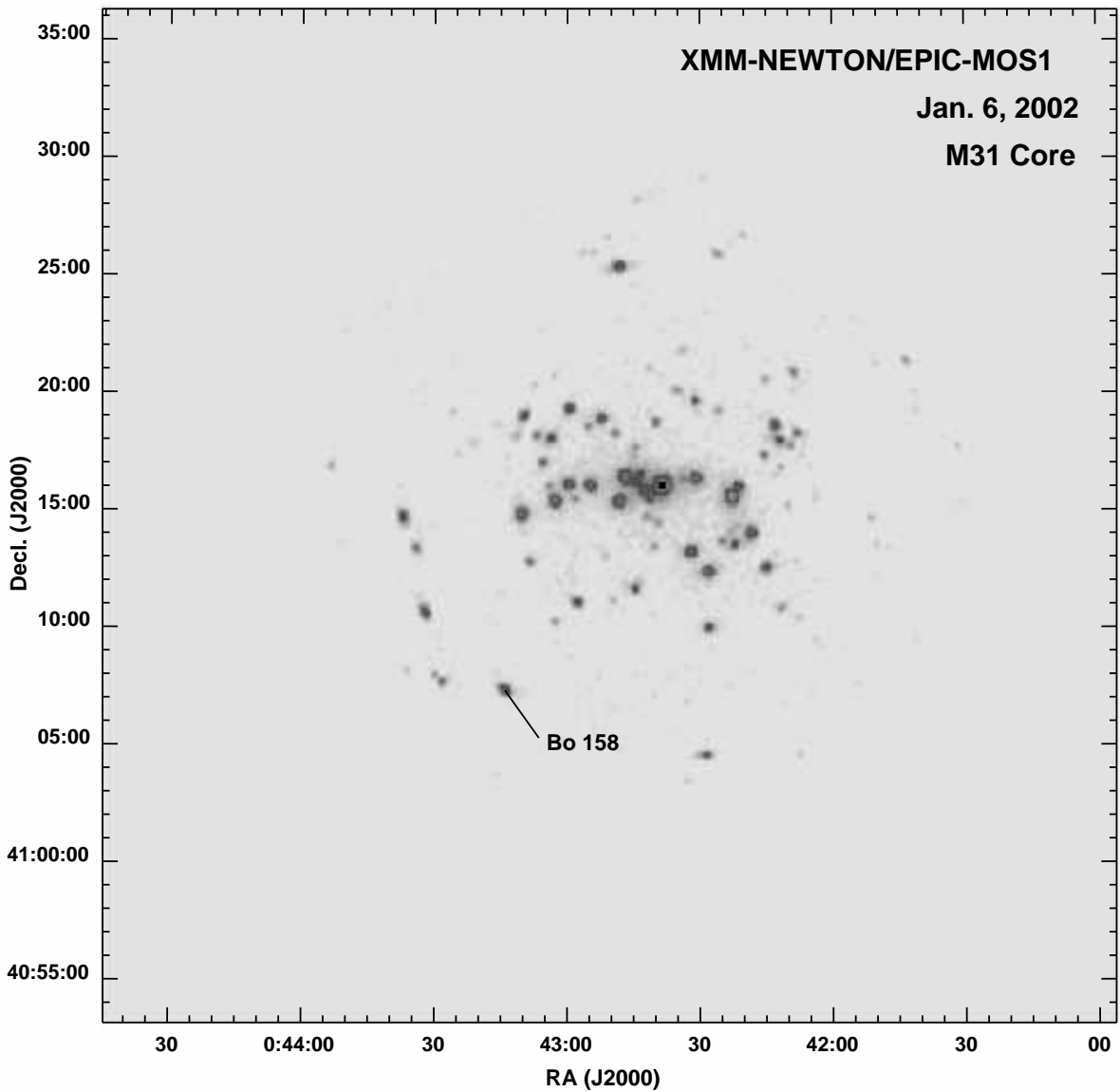}
\caption{\small X-ray image of the central part of M31, as it appears in 
the $63$-ks exposure with EPIC-MOS1 in the $0.3 - 10$ keV energy range. The 
position of the X-ray source XMMU J004314.1+410724 associated with globular 
cluster Bo158 (Battistini et al. 1987) is marked with an arrow. 
\label{image_mos}}
\end{figure}

In the following analysis we use data from three {\em XMM-Newton} observations 
of the bulge of M31 (Table \ref{obslog}, Figure \ref{image_mos}). The first 
observation of the central part of M31 was performed on June 25, 2000 as a 
part of the Performance Verification Program (PI: M.G. Watson) 
(\cite{Shirey01,Osborne01}). Two other observations were performed on June 
29, 2001 (\cite{Shirey01.1}) and on January 6, 2002 as a part of the 
Guaranteed Time Program (PI: K.O. Mason and M.G. Watson). We use data from 
three European Photon Imaging Camera (EPIC) instruments: two EPIC MOS 
detectors (\cite{Turner01}) and the EPIC PN detector (\cite{Strueder01}). In 
all observations the EPIC instruments were operated in the {\em full window 
mode} ($30\arcmin$ diameter FOV) with medium (2000 June 25 and 2001 June 29 
observations) and thin (2002 January 6 observation) optical blocking filters.

During all three {\em XMM-Newton} observations, the X-ray source Bo 158 was 
offset by $\sim 10\arcmin$ from the center of the field of view 
(Fig. \ref{image_mos}). In spite of the significant degradation 
in the sensitivity of the EPIC cameras at high offset angles, the statistics 
(more than 3000 counts in each observation) were sufficient for a detailed 
spectral and timing analysis.  

We reduced EPIC data with the {\em XMM-Newton} Science Analysis System (SAS 
v 5.3)\footnote{See http://xmm.vilspa.esa.es/user}. We performed standard 
screening of the EPIC data to exclude time intervals with high background 
levels. To generate lightcurves and spectra of the source, we used an 
extraction radius of $\sim 60 \arcsec$ and subtracted as background the 
spectrum of adjacent source-free regions with subsequent normalization by a 
ratio of detector areas. We used data in the $0.3 - 10$ keV energy band 
because of the uncertainties in the calibration of the EPIC instruments 
outside this range. All fluxes and luminosities presented below apply to 
this band. In the following analysis we assume a source distance of 760 
kpc (van den Bergh 2000).

We used standard XANADU/XRONOSv.5
\footnote{http://heasarc.gsfc.nasa.gov/docs/xanadu/xronos/xronos.html} 
Fourier transform and epoch folding tasks to search for the periodic 
modulation of the source X-ray flux and determine its period.

We used spectral response matrixes generated by SAS tasks. The energy spectra 
of the source were fitted to two analytic models using XSPEC v.11 
(\cite{Arnaud96}): an absorbed simple power law (powerlaw) and comptonization 
(comptt) models\footnote{For model description see 
http://heasarc.gsfc.nasa.gov/docs/xanadu/xspec/xspec.html and references therein}. 
EPIC-PN, MOS1 and MOS2 data were fitted simultaneously, but with independent 
normalizations.

\section{RESULTS}
\begin{figure}
\epsfxsize=18.0cm
\epsffile{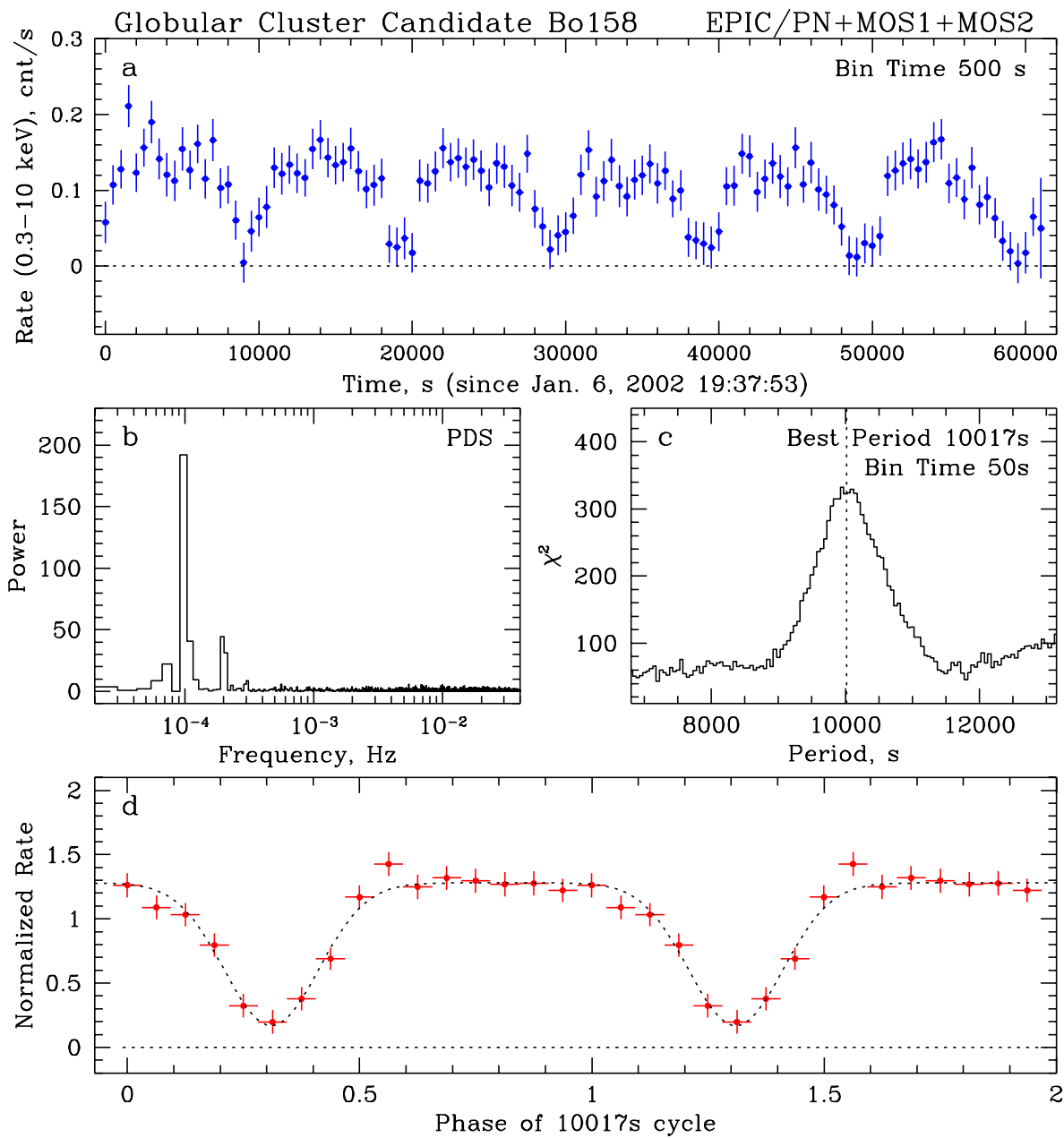}
\caption{\small ({\em a}) X-ray lightcurve of Bo158 during Jan. 6, 2002 
{\em XMM-Newton} observation, obtained from combined data of EPIC-PN, MOS1 
and MOS2 cameras, $0.3 - 10$ keV energy range, $500$ s time resolution. 
({\em b}) Power density spectrum of Bo 158. ({\em c}) Resulting $\chi^{2}$ 
distribution for the epoch folding analysis of the X-ray lightcurve. The 
best-fit value of the period (10017 s) is shown by a vertical dotted line. 
({\em d}) Normalized lightcurve of the source, folded at the best-fit period 
of $10017$ s. The analytic approximation is shown with a dotted line (see 
text). \label{timing_all}}
\end{figure}

\begin{figure}
\epsfxsize=18.0cm
\epsffile{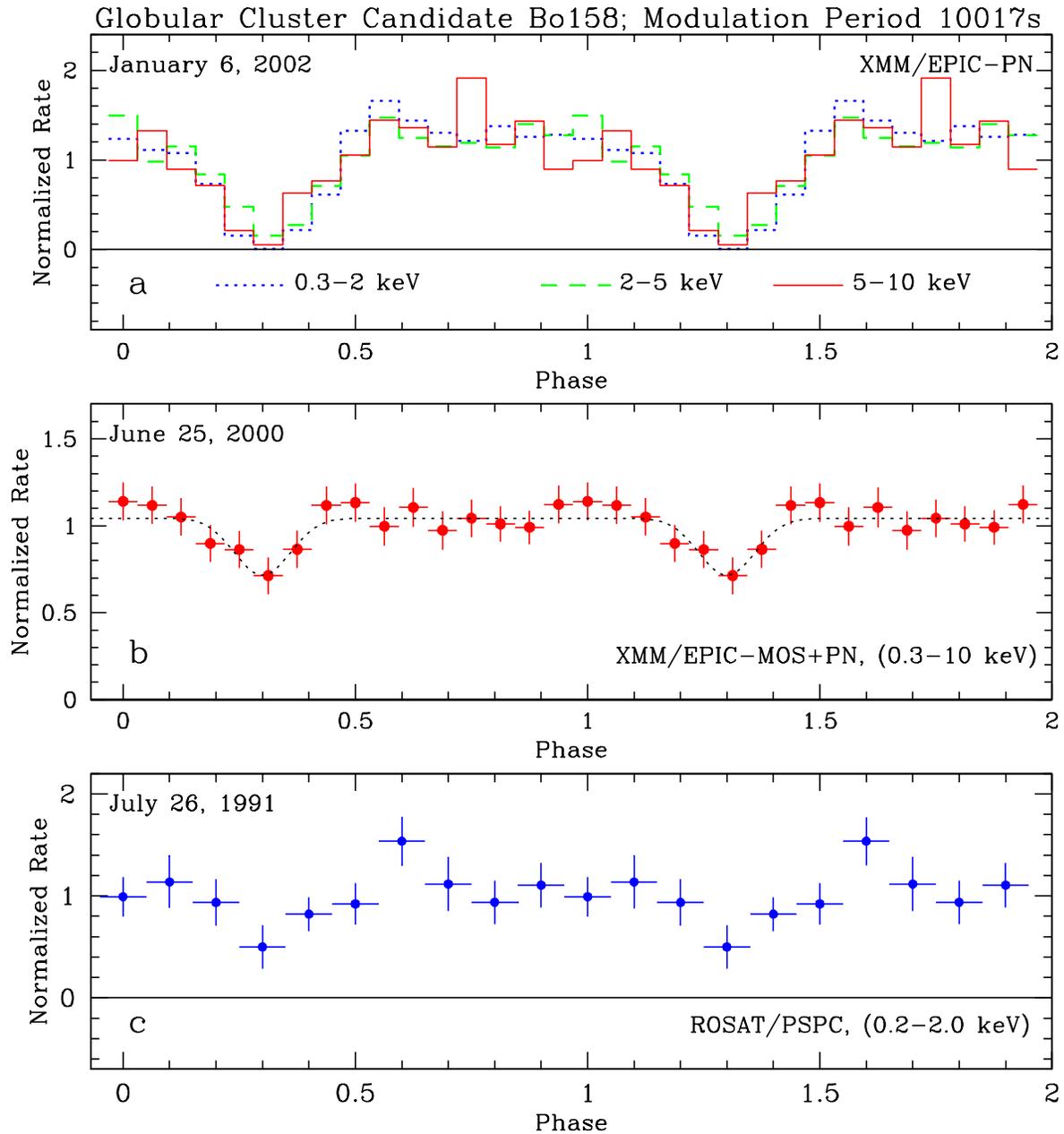}
\caption{\small X-ray lightcurves of Bo158 taken with 
{\em XMM-Newton}/EPIC and {\em ROSAT}/PSPC, folded on a period of 10017 s. 
({\em a}) Folded X-ray lightcurves of Bo158 in the $0.3 - 2$, $2 - 5$ keV 
and $5 - 10$ keV ranges during the January 6, 2002 {\em XMM-Newton} 
observation are shown with {\em dotted}, {\em long-dashed} and 
{\em thin solid} lines respectively (combined EPIC-MOS and PN data). 
({\em b}) Folded X-ray lightcurve of Bo158 in the $0.3 - 10$ keV energy 
range during the 2000 June 25 {\em XMM} observation (combined EPIC-MOS and 
PN data). The analytic approximation is shown with dotted line (see text). 
({\em c}) Folded X-ray lightcurve of Bo158 in the $0.2 - 2.0$ keV energy 
range during the 1991 June 26 {\em ROSAT}/PSPC observation. 
\label{timing_xmm_rosat}}
\end{figure}

We generated X-ray lightcurves of the source in the $0.3 - 10$ keV energy 
band using data from the EPIC-PN, MOS1 and MOS2 detectors, combining them in 
order to improve statistics. The resulting lightcurve of Bo 158 for the 
2002 Jan. 6 observation (Fig. \ref{timing_all}{\em a}) shows a pattern of 
recurrent dips in the X-ray intensity with a period of $\sim 10000$ s. 
The dips are broad, lasting for $\sim 30\%$ of the $10000$-s cycle. 
Their FWHM varies between 2200 and 3000 s, and their depths range from 
$\sim 80$ to $\sim 100\%$ of the out-of-dip flux. The recurrence period is 
$10017\pm50$ s. Figure \ref{timing_all}{\em b,c} shows the power density 
spectrum of the source from $2\times 10^{-5}$ to $3\times 10^{-2}$ Hz, and 
our determination of the period with an epoch folding analysis. Figure 
\ref{timing_all}{\em d} shows the combined EPIC (PN+MOS1+MOS2) light curve 
for the Jan. 6 observation folded on the 10017 s best period of the 
modulation (phase 0.0 is set arbitrarily at TJD=12279.0). In order to 
investigate the energy dependence of the X-ray modulation during 2002 Jan. 6 
observation, we constructed lightcurves in the $0.3 - 2.0$, $2.0 - 5.0$, 
and $5.0 -10.0$ keV bands. As it is clearly seen in Figure 
\ref{timing_xmm_rosat}{\em a}, there is no significant energy dependence of 
the modulation.

We searched for this modulation in earlier observations of M31, including 
2000 Jun. 25 and 2001 Jun. 29 {\em XMM-Newton} observations of the center 
of M31, and {\em ROSAT}/PSPC observation from 1991 Jul. 26. We produced 
combined EPIC-MOS lightcurves in the $0.3 - 10.0$ keV energy range and a 
PSPC lightcurve in the $0.2 - 2.0$ keV energy band. X-ray flux modulations 
with periods close to $\sim 10000$ s ($\sim 10500$ s and $\sim 9700$ s 
respectively) were marginally detected in the 2000 Jun. 25 and 1991 Jul. 26 
observations. Figure \ref{timing_xmm_rosat} shows the {\em XMM-Newton}/EPIC 
and {\em ROSAT}/PSPC light curves folded on the 10017 s period determined 
above. The lightcurves of 2000 Jun. 25 and 1991 Jun. 26 observations show a 
$\sim 30\%$ and $\sim 50\%$ intensity drop during the dip (Fig. 
\ref{timing_xmm_rosat}{\em b,c}), while for the 2001 Jun. 29 observation, 
dips were not detected with a $2\sigma$ upper limit of $10\%$.

In case of 2002 Jan. 6 and 2000 Jun. 25 observations, where regular dips 
in the X-ray lightcurve of the source were clearly detected, we fit the 
folded $0.3 - 10$ keV lightcurves with a simple model, consisting of a 
constant plus a Gaussian with negative normalization centered at the dip 
minimum: $f(p) = C - A \times exp [-(p-p_{\rm dip})^{2}/2 \sigma^{2})]$
\footnote{where $p$ is a phase of 10017-s cycle, $C$ is a source 
out-of-dip intensity level, $A$, $p_{dip}$ and $2.35\sigma$ are 
normalization, centroid phase and FWHM of Gaussian respectively}(the 
best-fit model approximation is shown with dotted lines in 
Fig. \ref{timing_all}{\em d} and Fig. \ref{timing_xmm_rosat}{\em b}). Using 
this approximation, we obtain a $83\pm5\%$ amplitude of the dip with respect 
to the out-of-dip intensity and $\sigma = 0.10\pm0.01$ (in phase units of 
$10017$ s cycle) for the 2002 Jan. 6 observation, and $30\pm9\%$ amplitude 
and $\sigma = 0.07\pm0.03$ for the 2000 Jun. 25 observation. The values of 
$\sigma$ imply an average FWHM for the dips of order $2500$ s.

The averaged {\em XMM}/EPIC spectra of Bo 158 was analyzed by fitting 
two different spectral models (Table \ref{spec_par}). A model with 
a quasi-exponential cut-off (the Comptonization model in Table \ref{spec_par}) 
at $\sim 4.0 - 6.0$ keV describes the energy spectra significantly better 
than a simple power law. This spectrum is somewhat reminiscent of Galactic 
neutron star systems with high luminosity (\cite{Iaria01,DiSalvo01}). The 
spectrum is stable from observation to observation, despite the change in 
X-ray luminosity from $\sim 6.5 \times 10^{37}$ (2002 Jan. 6) to 
$\sim 1.4 \times 10^{38}$ ergs s$^{-1}$ (2001 Jun. 29), although the cut-off 
energy may increase at the higher luminosities.

\begin{figure}
\epsfxsize=18.0cm
\epsffile{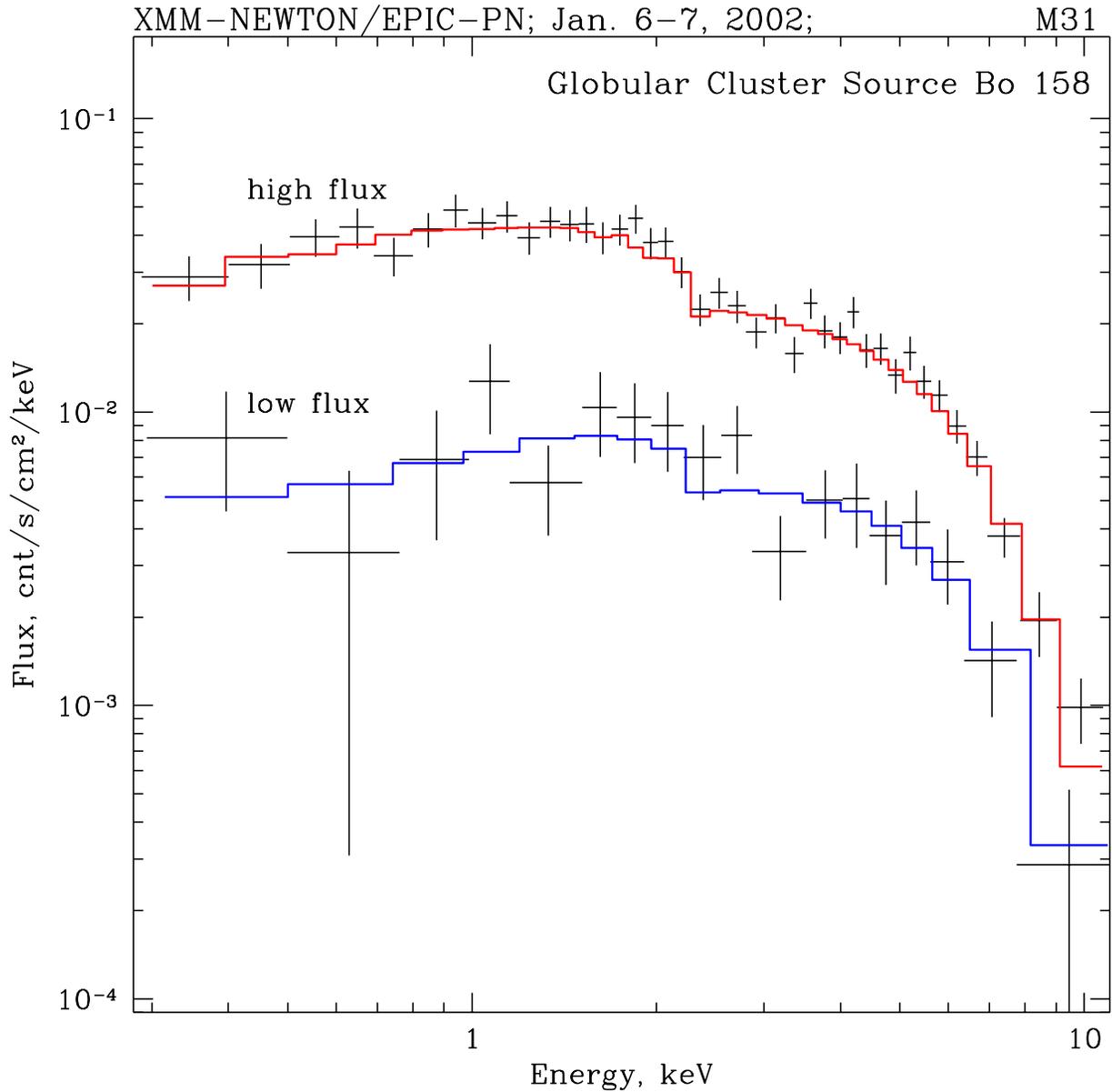}
\caption{\small The energy spectra of Bo158 from the non-dip (high 
flux) and dip (low flux) intervals of X-ray flux during the 2002 Jan. 6 
observation. EPIC-PN data in the $0.3 - 10$ keV energy range. The 
best-fit analytic models are shown with {\em red} and {\em blue} histograms, 
and are both saturated comptonization models with an electron temperature of 
$\sim 1.7$ keV and an optical depth of $\sim 22$. \label{spectra_high_low}}
\end{figure}

In order to investigate any changes in the energy spectrum during the 
intensity dips, we divide the spectra from 2002 Jan. 6 observation into 
two states corresponding to the out-of-dip and dip intensity 
(Fig. \ref{spectra_high_low})(i.e. the fluxes higher and lower than $0.1$ 
cnt/s in Fig. \ref{timing_all}{\em a}). A saturated comptonization model 
with electron temperature $kT_{e} \sim 1.7$ keV and optical depth 
$\tau \sim 22$ gives a good approximation to both spectra 
(Fig. \ref{spectra_high_low}). The resulting $2\sigma$ upper limit 
on an increase in neutral absorbing column density $N_{\rm H}$ during 
the dips is $2 \times 10^{21}$ (assuming a single spectral component). 
However, if the dips are caused by electron scattering in a cloud along 
the line of sight, the electron column density must be 
$\sim 2.7 \times 10^{24}$ cm$^{-2}$ ($83\%$ dipping) and 
$\sim 5.4 \times 10^{23}$ cm$^{-2}$ ($30\%$ dipping) for 2002 Jan. 6 and 
2000 Jun. 25 observations respectively.

It should be mentioned that the Comptonization model which fits the spectrum 
during the intensity dips does not provide a unique determination of the 
spectral form. Other complex models (e.g. two-component models with different 
low-energy absorption column densities) also give acceptable fits. The low 
statistical significance of the data does not allow us to discriminate 
between these models.

There is an anti-correlation between the average amplitude of the 
modulation and the source intensity on a long time scale. The values of 
modulation fraction were $83\%$ in 2002 Jan. 6 observation, $30\%$ in 
2000 Jun. 25 observation and $<10\%$ in 2001 Jun. 29 observation, 
corresponding to the source luminosities of $6.5 \times 10^{37}$, 
$9.9 \times 10^{37}$ and $1.4 \times 10^{38}$ ergs s$^{-1}$ respectively. 
In addition, a dependence of the dip strength upon the luminosity 
on a time scale of hours was marginally detected during the 2000 Jun. 25 
observation.

\section{DISCUSSION}
We have discovered a $2.78$-hour periodic dip-like modulation with 
variable amplitude in the X-ray flux of the globular cluster candidate 
source Bo 158 in M31. The amplitude of the modulation is anticorrelated 
with source intensity: it becomes lower with increasing X-ray flux. The 
most interesting feature of this modulation is its lack of energy 
dependence. 

If the $2.78$-hour modulation represents the binary period, $P_{\rm orb}$, 
the binary separation is $a \sim 7 \times 10^{10} M_{\rm X}^{1/3} (1+q)^{1/3}$ 
cm \footnote{where $M_{\rm X}$ is a mass of the compact object in solar 
units and $q$ is a mass ratio of the secondary star and a compact object}. 
The association of the source with a globular cluster, and spectral
parameters consistent with Galactic neutron star sources, suggest that the 
compact object in Bo158 is probably a neutron star in a low-mass binary. This 
implies a highly compact binary with separation $a \lesssim 10^{11}$ cm.

The properties of Bo 158 are somewhat reminiscent of Galactic X-ray 
sources that exhibit dip-like modulations (e.g. Cygnus X-3 and some "dipping" 
sources). 

Although there are similarities between the light curve of Bo 158 and the 
Galactic binary Cygnus X-3, it is unlikely that these systems have a common 
nature. The broad minimum in the 4.8-hour cycle of Cygnus X-3 is explained 
as variable scattering in the dense photoionized wind of a Wolf-Rayet 
companion (\cite{Paerels00}). Moreover, in contrast to Bo 158, the 
amplitude of X-ray modulation in Cygnus X-3 is extremely stable in spite 
of drastic changes in luminosity.

The duration of the intensity dips in phase is similar to those seen in 
some Galactic LMXB dipping sources such as X1755-338 (\cite{Mason85}), 
XB1254-690 (\cite{Courvoisier86}) and XB 1916-053 (\cite{Yoshida95}), 
although Bo 158 is much more luminous. The lack of energy dependence of the 
modulation is a common feature of a number of dipping sources. The dips 
in these systems are believed to be caused by absorption and scattering in 
an obscuring medium, i.e. a bulge or thickened region of the accretion disk, 
or inhomogeneities in the mass transfer stream (\cite{ws82,LSHU76,FKL87}). 

If the dips in Bo 158 are caused by an obscuring structure, at least two 
explanations could be proposed to explain the energy independence of the 
intensity dips: i) the obscuring material is highly ionized (\cite{Mason85}); 
ii) the dips are caused by partial covering of an extended source 
(\cite{Church97}). 

(i) If the obscuring structure is located close to the X-ray emission region, 
the obscuring medium could be strongly ionized. The compactness and high 
X-ray luminosity of Bo 158 could imply a high ionization in the region 
responsible for the dips. The obscuring matter in the Galactic dipping 
sources, with their smaller luminosities ($10^{36} - 10^{37}$ ergs s$^{-1}$), 
is less strongly ionized. For the material causing the dips to be completely 
ionized, the ionization parameter 
$\xi = L_{\rm X}/n R^{2}$ (where $L_{\rm X}$ is the central source luminosity, 
$n$ is the gas density of the cloud, and $R$ is the distance from the central 
source to the obscuring medium)(\cite{HBM76}) should be larger than 
$\sim 1000$ ergs cm s$^{-1}$. For the 2002 Jan. 6 observation, an average 
$83 \%$ intensity reduction requires a column, $N_{\rm e}$, of 
$\sim 2.7 \times 10^{24}$ cm$^{-2}$. Assuming an out-of-dip luminosity of 
Bo 158 to be $6.5 \times 10^{37}$ ergs s$^{-1}$, and following Mason et al. 
(1985), we conclude that all material closer than 
$\sim 2.4 \times 10^{10} \epsilon$ cm \footnote{where filling factor, 
$\epsilon = (l/R)$, $l$ is the scattering column length} will be highly 
ionized. Another estimate of the ionization stage of the obscuring medium is 
based on the duration of the intensity dips (\cite{rc84}). For $t_{\rm dip} 
\sim 2500$ s, assuming that the absorbing medium is roughly spherical and 
fixed in the frame of the binary, 
$\xi = 2 \pi L_{\rm X} t_{\rm dip}/(N P_{\rm orb} R)$. To obtain 
$\xi \gtrsim 1000$, the ionized region must be closer than 
$R \sim 4 \times 10^{10}$ cm (2002 Jan. 6 observation) and 
$R \sim 2 \times 10^{11}$ cm (2000 Jun. 25 observation). The absorbing 
medium in a 2.78-hour binary could fall well within these limits, and 
be highly ionized.

(ii) For an opaque absorber to produce broad dips, the emission region 
must be extended and the eclipse partial. The source could be a scattering 
corona-like structure, and/or thick disk/outflow located close to the 
compact object, and the absorber a thickened region of accretion disk, or 
a mass transfer stream from the companion star. 

The diminution of the modulation with increase of the X-ray flux could 
indicate a decrease of the obscured fraction of the emission region, caused 
by increasing the size of the emitter or decreasing the size of the absorber. 

We would like to thank the referee for his/her helpful comments. This paper 
is based in part on observations obtained with {\em XMM-Newton}, an ESA 
science mission with instruments and contributions directly funded by ESA 
Member States and the USA (NASA). This research has made use of data obtained 
through the High Energy Astrophysics Science Archive Research Center Online 
Service, provided by the NASA Goddard Space Flight Center.

\clearpage

\begin{table}
\small
\caption{XMM-Newton observations of M31 used in this analysis. 
\label{obslog}}
\tiny
\begin{tabular}{cccccccc}
\hline
\hline
Date, UT & $T_{\rm start}$, UT & Field & Obs. ID  & RA (J2000)$^{a}$ & Dec (J2000)$^{a}$ & Exp.(MOS)$^{b}$ & Exp.(PN)$^{b}$\\
&(h:m:s)&&&  (h:m:s)   &(d:m:s)&(ks)&(ks)\\             
\hline
25/06/2000 &10:44:42&M31 Core  &0112570401&00:42:43.0&41:15:46.1&34.8&31.0\\
29/06/2001 &06:21:38&M31 Core  &0109270101&00:42:43.0&41:15:46.1&32.6&30.8\\
06/01/2002 &18:07:17&M31 Core  &0112570101&00:42:43.0&41:15:46.1&63.0&61.0\\
\hline
\end{tabular}
\begin{list}{}{}
\item[$^{a}$] -- coordinates of the center of the FOV
\item[$^{b}$] -- instrument exposure used in the analysis 
\end{list}
\end{table}

\begin{table}
\small
\caption{Best-fit model parameters of the energy spectra of Bo 158 during 
June 25, 2000, June 29, 2001 and January 6, 2002 {\em XMM - Newton} 
observations of the central part of M31 (combined EPIC-PN, MOS1 and MOS2 
data). Parameter errors correspond to $1 \sigma$ level.
\label{spec_par}}
\small
\begin{tabular}{cccc}
\hline
\hline
Parameter           & \multicolumn{3}{c}{Observation Date}\\
                    & Jun. 25, 2000 & Jun. 29, 2001 & Jan. 6, 2002\\
\hline
\multicolumn{4}{c}{Absorbed Power Law (powerlaw*wabs)}\\
\hline
Photon Index         &$0.66^{+0.04}_{-0.03}$&$0.58^{+0.04}_{-0.02}$&$0.66\pm0.03$\\
N$_{\rm H}^{\rm a}$  &$0.09\pm0.02$         &$0.05$&$0.07^{+0.02}_{-0.01}$\\
Flux$^{\rm b}$       &$1.738\pm0.035$&$2.401\pm0.031$&$1.082\pm0.019$\\
$\chi^{2}$(d.o.f)    &$302.3(262)$&$414.6(355)$&$287.4(267)$\\
\hline
\multicolumn{4}{c}{Absorbed Comptonization Model (comptt*wabs)}\\
\hline
kT$_{0}^{\rm c}$     &$0.08_{-0.07}^{+0.05}$&$0.08^{+0.05}_{-0.07}$&$0.09^{+0.08}_{-0.07}$\\
kT$_{\rm e}^{\rm d}$ &$1.78^{+0.10}_{-0.09}$&$1.87^{+0.08}_{-0.06}$&$1.73\pm0.07$\\
$\tau^{\rm e}$       &$19.7^{+2.3}_{-1.3}$  &$20.1^{+2.1}_{-1.0}$  &$21.6^{+2.2}_{-1.2}$\\
N$_{\rm H}^{\rm a}$  &$0.08^{+0.02}_{-0.03}$&$0.05^{+0.02}_{-0.05}$&$0.05^{+0.02}_{-0.01}$\\
Flux$^{b}$           &$1.442\pm0.029$       &$2.005\pm0.026$       &$0.914\pm0.016$\\
$\chi^{2}$(d.o.f)    &$263.8(260)$          &$358.1(353)$          &$194.4(265)$\\
\hline
\end{tabular}

\begin{list}{}{}
\item $^{\rm a}$ -- equivalent absorbing hydrogen column density in units of $10^{22}$ cm$^{-2}$
\item $^{\rm b}$ -- model flux in the $0.3 - 10.0$ keV energy range in units of $10^{-12}$ erg s$^{-1}$ cm$^{-2}$
\item $^{\rm c}$ -- temperature of soft photons, keV
\item $^{\rm d}$ -- electron temperature, keV
\item $^{\rm e}$ -- Thomson optical depth for the spherical geometry
\end{list}

\end{table}

\clearpage

\clearpage

\clearpage

\clearpage

\end{document}